\def\aa{{A\&A}}

\def\aj{{AJ}}

\def\apj{{ApJ}}
\def\apjs{{ApJS}}

\def\mnras{{MNRAS}}

\def\prd{{Phys. Rev. D}}
\newcommand{\be}{\begin{equation}}
\newcommand{\ee}{\end{equation}}
\newcommand{\bea}{\begin{eqnarray}}
\newcommand{\eea}{\end{eqnarray}}

\newcommand{\M}{{\rm M}}
\newcommand{\R}{{\rm R}}
\newcommand{\vac}{{\rm vac}}
\newcommand{\dark}{{\rm dark}}

\newcommand{\susy}{{\rm SUSY}}

\documentclass[cup5b]{caps}
\usepackage{graphicx}
\usepackage{amssymb}
\usepackage{ociwsymp2}   
\HeadText{Sean M. Carroll}

\begin{document}

\pagenumbering{arabic}

\author[]{SEAN M. CARROLL
\\Enrico Fermi Institute, Department of Physics, and Center
for Cosmological Physics,
\\University of Chicago
\\Kavli Institute for Theoretical Physics, University of
California, Santa Barbara
\\carroll@theory.uchicago.edu}

\chapter{Why is the Universe Accelerating?}

\begin{abstract}

The universe appears to be accelerating, but the reason why is 
a complete mystery.  The simplest explanation, a small vacuum energy
(cosmological constant), raises three difficult issues:  
why the vacuum energy is so small, why it is not quite
zero, and why it is comparable to the matter density today.
I discuss these mysteries, some of their possible resolutions, and
some issues confronting future observations.

\end{abstract}

\section{Introduction}

Recent astronomical observations have provided strong evidence that
we live in an accelerating universe.  By itself, acceleration is
easy to understand in the context of general relativity and quantum 
field theory; however, the very small but nonzero energy scale
seemingly implied by the observations is completely
perplexing.  In trying to understand the universe in which we apparently
live, we are faced with a problem, a puzzle, and a scandal:
\begin{itemize}
\item The {\bf cosmological constant problem:}  why is the energy of
the vacuum so much smaller than we estimate it should be?
\item The {\bf dark energy\footnote{``Dark
energy'' is not, strictly speaking, the most descriptive name for
this substance; lots of things are dark, and everything has energy.
The feature which distinguishes dark energy from ordinary matter
is not the energy but the pressure, so ``dark pressure'' would be
a better term.  However, it is not the existence of the pressure,
but the fact that it is negative -- tension rather than ordinary
pressure -- that drives the acceleration of the universe, so 
``dark tension'' would be better yet.  And we would have detected 
it long ago if it had collected into potential
wells rather than being smoothly distributed, so ``smooth tension''
would be the best term of all, not to mention sexier.  I thank Evalyn
Gates, John Beacom, and Timothy Ferris for conversations on this
important point.} puzzle:}  what is the nature of the
smoothly-distributed, persistent energy density which appears to 
dominate the universe?
\item The {\bf coincidence scandal:}  why is the dark energy density
approximately equal to the matter density today?
\end{itemize}
Any one of these issues would represent a serious challenge to 
physicists and astronomers; taken together, they serve to remind us
how far away we are from understanding one of the most basic features
of the universe.

The goal of this article is to present
a pedagogical (and necessarily superficial) introduction to the 
physics issues underlying these questions, rather than a comprehensive
review; for more details and different points of view see
Sahni and Starobinski (2000), Carroll (2001), Padmanabhan (2003), 
or Peebles and Ratra (2003).
After a short discussion of the issues just mentioned, we will turn
to mechanisms which might address any or all of them; we will pay special
attention to the dark energy puzzle, only because there is more to
say about that issue than the others.  We will close with an
idiosyncratic discussion of issues confronting observers studying
dark energy.

\section{The mysteries}

\subsection{Classical vacuum energy}

Let us turn first to the issue of why the vacuum energy is
smaller than we might expect.  When Einstein proposed general
relativity, his field equation was
\begin{equation}
  R_{\mu\nu} - {1\over 2}Rg_{\mu\nu} 
  = 8\pi GT_{\mu\nu}\ ,
  \label{einstein}
\end{equation}
where the left-hand side characterizes the geometry of spacetime
and the right-hand side the energy sources; $g_{\mu\nu}$ is the
spacetime metric, $R_{\mu\nu}$ is the Ricci tensor, $R$ is the
curvature scalar, and $T_{\mu\nu}$ is the energy-momentum tensor.
(I use conventions in which $c=\hbar=1$.)  If the energy sources
are a combination of matter and radiation, there are no solutions
to (\ref{einstein}) describing a static, homogeneous universe.
Since astronomers at the time believed the universe was static,
Einstein suggested modifying the left-hand side of his equation
to obtain
\begin{equation}
  R_{\mu\nu} - {1\over 2}Rg_{\mu\nu} 
  + \Lambda g_{\mu\nu}
  = 8\pi GT_{\mu\nu}\ , 
  \label{einsteinl}
\end{equation}
where $\Lambda$ is a new free parameter, the cosmological 
constant.  This new equation admits a static, homogeneous solution
for which $\Lambda$, the matter density, and the spatial curvature 
are all positive:  the ``Einstein static universe.''  The need for
such a universe was soon swept away by improved astronomical 
observations, and the cosmological constant acquired a somewhat
compromised reputation.

Later, particle physicists began to contemplate the possibility 
of an energy density inherent in the vacuum (defined as the 
state of lowest attainable energy).  If the vacuum is to look
Lorentz-invariant to a local observer, its 
energy-momentum tensor must take on the unique form
\begin{equation}
  T^{\rm vac}_{\mu\nu} = -\rho_{\rm vac} g_{\mu\nu}\ ,
  \label{tmunuvac}
\end{equation}
where $\rho_{\rm vac}$ is a constant vacuum energy density.
Such an energy is associated with an isotropic pressure
\begin{equation}
  p_{\rm vac} = -\rho_{\rm vac}\ .
  \label{pemr}
\end{equation}
Comparing this kind of energy-momentum tensor to the appearance of the
cosmological constant in (\ref{einsteinl}), we find that they
are formally equivalent, as can be
seen by moving the $\Lambda g_{\mu\nu}$ term in
(\ref{einsteinl}) to the right-hand side and setting
\begin{equation}
  \rho_{\rm vac} = \rho_\Lambda \equiv {{\Lambda}\over{8\pi G}}\ .
\end{equation}
This equivalence is the origin of the identification of
the cosmological constant with the energy of the vacuum.

From either side of Einstein's equation, the 
cosmological constant $\Lambda$ is 
a completely free parameter.  It has dimensions of [length]$^{-2}$
(while the energy density $\rho_\Lambda$ has units [energy/volume]),
and hence defines a scale, while general
relativity is otherwise scale-free.  Indeed, from purely classical
considerations, we can't even say whether
a specific value of $\Lambda$ is ``large'' or ``small''; it is simply
a constant of nature we should go out and determine through
experiment. 

\subsection{Quantum zero-point energy}

The introduction of quantum mechanics changes this story somewhat.
For one thing, Planck's constant allows us to define a gravitational
length scale, the reduced Planck length
\be
  L_{\rm P} = \left({8\pi G}\right)^{1/2}
  \sim 10^{-32}~{\rm cm}
\ee
as well as the reduced Planck mass
\be
  M_{\rm P} = \left({1\over 8\pi G}\right)^{1/2}
  \sim 10^{18}~{\rm GeV}\ ,
\ee
where ``reduced'' means that we have included the $8\pi$'s where they
really should be.  (Note that, with $\hbar=1$ and $c=1$, we have
$L=T=M^{-1}=E^{-1}$, where $L$ represents a length scale, 
$T$ a time interval, $M$
a mass scale, and $E$ an energy.)  Hence, 
there is a natural expectation for the scale of the cosmological
constant, namely
\be
  \Lambda^{\rm (guess)} \sim L_{\rm P}^{-2}\ ,
\ee
or, phrased as an energy density,
\be
  \rho_{\rm vac}^{\rm (guess)} \sim M_{\rm P}^4 
  \sim (10^{18}{\rm ~GeV})^4 \sim 10^{112} {\rm ~erg/cm}^3\ .
  \label{rhoguess}
\ee

We can partially justify this guess by thinking about 
quantum fluctuations in the vacuum.  At all energies probed by 
experiment to date, the world is accurately described as a set
of quantum fields (at higher energies it may become strings or
something else).  If we take the Fourier transform
of a free quantum field, each mode of fixed wavelength 
behaves like a simple
harmonic oscillator.  (``Free'' means ``noninteracting''; for our
purposes this is a very good approximation.)  As we know from 
elementary quantum mechanics, the ground-state or zero-point energy
of an harmonic oscillator with potential $V(x)={1\over 2}\omega^2 x^2$
is $E_0 = {1\over 2}\hbar \omega$.  Thus, each mode of a quantum
field contributes to the vacuum energy, and the net result should be
an integral over all of the modes.  Unfortunately this integral
diverges, so the vacuum energy appears to be infinite.
However, the infinity arises from the contribution of modes with
very small wavelengths; perhaps it was a mistake to include such modes,
since we don't really know what might happen at such scales.
To account for our ignorance, we could introduce a cutoff energy,
above which we ignore any potential contributions, and
hope that a more complete theory will eventually provide a physical 
justification for doing so.  If this cutoff is at the Planck scale,
we recover the estimate (\ref{rhoguess}).

The strategy of decomposing a free field into individual modes and
assigning a zero-point energy to each one really only makes sense
in a flat spacetime background.  In curved spacetime we can still
``renormalize'' the vacuum energy, relating the classical parameter
to the quantum value by an infinite constant.  After renormalization,
the vacuum energy is completely arbitrary, just as it was in the original
classical theory.  But when we use general relativity we are really
using an effective field theory to describe a certain limit of quantum
gravity.  In the context of effective field theory, if a parameter
has dimensions [mass]$^n$, we expect the corresponding mass parameter
to be driven up to the scale at which the effective description
breaks down.  Hence, if we believe classical
general relativity up to the Planck scale, we would expect the
vacuum energy to be given by our original guess (\ref{rhoguess}).

However, we believe we have now measured the vacuum energy through
a combination of Type~Ia supernovae (Riess et al.\ 1998, Perlmutter
et al.\ 1999, Tonry et al.\ 2003, Knop et al.\ 2003), 
microwave background anisotropies (Spergel et al.\ 2003),
and dynamical matter measurements (Verde et al.\ 2003), to reveal
\be
  \rho_{\rm vac}^{\rm (obs)} \sim  10^{-8}{\rm ~erg/cm}^3
  \sim (10^{-3}{\rm ~eV})^4\ ,
  \label{vescale}
\ee
or
\be
  \rho_{\rm vac}^{\rm (obs)}
  \sim  10^{-120}\rho_{\rm vac}^{\rm (guess)}\ . 
  \label{rhoobs}
\ee
For reviews see Sahni and Starobinski 2000, Carroll 2001, or 
Peebles and Ratra 2003.

Clearly, our guess was not very good.
This is the famous 120-orders-of-magnitude discrepancy that makes
the cosmological constant problem such a glaring embarrassment.
Of course, it is somewhat unfair to emphasize the factor of
$10^{120}$, which depends on the fact that energy density
has units of [energy]$^4$.  We can express the vacuum energy in
terms of a mass scale,
\be
  \rho_\vac = M_\vac^4\ ,
\ee
so our observational result is
\be
  M_\vac^{{\rm (obs)}} \sim 10^{-3}{\rm ~eV}\ .
\ee
The discrepancy is thus
\be
  M_\vac^{{\rm (obs)}} \sim 10^{-30} M_\vac^{{\rm (guess)}}\ .
  \label{mvac1}
\ee
We should think of the cosmological constant problem as a
discrepancy of 30 orders of magnitude in energy scale.  

\subsection{The coincidence scandal}

The third issue mentioned above is the coincidence between the
observed vacuum energy (\ref{rhoobs}) and the current matter
density.  To understand this, we briefly review the dynamics of
an expanding Robertson-Walker spacetime.
The evolution of a homogeneous and isotropic universe is
governed by the Friedmann equation,
\be
  H^2 = {8\pi G\over 3}\rho - {\kappa \over a^2}\ ,
  \label{feq}
\ee
where $a(t)$ is the scale factor, $H=\dot{a}/a$ is the Hubble 
parameter, $\rho$ is the energy density, and $\kappa$ is the
spatial curvature parameter.  The energy density is a sum of
different components, $\rho = \sum_i\rho_i$, which will in 
general evolve differently as the universe expands.  For matter
(non-relativistic particles) the energy density goes as
$\rho_\M \propto a^{-3}$, as the number density is diluted with
the expansion of the universe.  For radiation the energy density
goes as $\rho_\R \propto a^{-4}$, since each particle loses energy
as it redshifts in addition to the decrease in number density.
Vacuum energy, meanwhile, is constant throughout spacetime, so
that $\rho_\Lambda \propto a^0$.

It is convenient to characterize the energy density
of each component by its density parameter 
\be
  \Omega_i = {\rho_i \over \rho_c}\ ,
\ee
where the critical density
\be
  \rho_c = {3H^2 \over 8\pi G}\ 
\ee
is that required to make the spatial geometry of the universe be
flat ($\kappa=0$).  The ``best-fit universe" or
``concordance" model implied by numerous observations
includes radiation, matter, and vacuum energy, with
\bea
\Omega_{\R 0} &\approx& 5\times 10^{-5}\cr
\Omega_{\M 0} &\approx& 0.3\cr
\Omega_{\Lambda 0} &\approx& 0.7\ ,
\eea 
together implying a flat universe.  We see that the densities in matter
and vacuum are of the same order of magnitude.\footnote{Of course
the ``matter'' contribution consists both of ordinary baryonic
matter and non-baryonic dark matter, with $\Omega_{\rm b}\approx
0.04$ and $\Omega_{\rm DM}\approx 0.25$.  The similarity between 
these apparently-independent quantities is another coincidence
problem, but at least one which is independent of time; we have 
nothing to say about it here.}  But the ratio of these
quantities changes rapidly as the universe expands:
\be
  {\Omega_\Lambda \over \Omega_\M} = {\rho_\Lambda
  \over \rho_\M} \propto a^3\ .
\ee
As a consequence,
\begin{figure}[t]
  \centering
  \includegraphics[height=10cm]{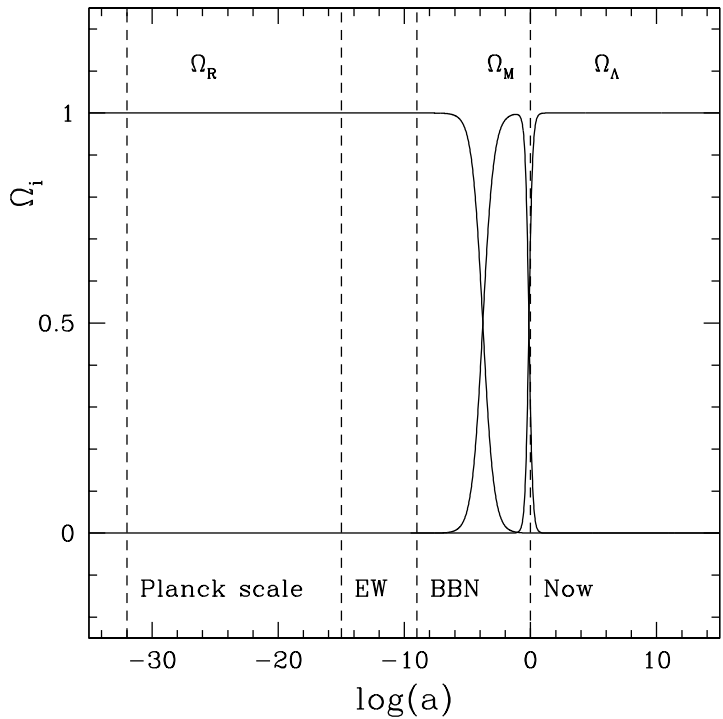}
  \caption{Density parameters $\Omega_i$ for radiation (R),
  matter (M), and vacuum ($\Lambda$), 
  as a function of the scale factor $a$, in
  a universe with $\Omega_{\Lambda 0} = 0.7$, $\Omega_{\M 0} = 0.3$,
  $\Omega_{\R 0} = 5\times 10^{-5}$.
  Scale factors corresponding to the Planck era, electroweak symmetry
  breaking (EW), and Big Bang nucleosynthesis (BBN) are indicated, as
  well as the present day.}
  \label{oplotall}
\end{figure}
at early times the vacuum energy was negligible in comparison to
matter and radiation, while at late times matter and radiation are
negligible.  There is only a brief epoch of the universe's history
during which it would be possible to
witness the transition from domination by
one type of component to another.  This is illustrated in 
Figure~\ref{oplotall}, in which the various density parameters
$\Omega_i$ are plotted as a function of the scale factor.
At early times $\Omega_{\rm R}$ is close to unity; the matter-radiation
transition happens relatively gradually, while the matter-vacuum
transition happens quite rapidly.  

How finely-tuned is it that we exist in the era when vacuum and
matter are comparable?  Between the Planck time and now, the universe
has expanded by a factor of approximately $10^{32}$.  To be fair,
we should consider an interval of logarithmic expansion which is
centered around the present time; this would describe a total expansion
by a factor of $10^{64}$.  If we take the transitional period 
between matter and vacuum to include the time from 
$\Omega_\Lambda/\Omega_\M = 0.1$ to $\Omega_\Lambda/\Omega_\M = 10$, the universe expands by a factor
of $100^{1/3} \approx 10^{0.67}$.  Thus, there is an
approximately $1\%$ chance that an
observer living in a randomly selected logarithmic expansion interval
in the history of our universe would be lucky enough to have
$\Omega_\M$ and $\Omega_\Lambda$ be the same order of magnitude.
Everyone will have their own favorite way of quantifying such
unnaturalness, but the calculation here
gives some idea of the fine-tuning involved; it is substantial,
but not completely ridiculous.

As we will discuss below, there is room to imagine that we are
actually not observing the effects of an ordinary cosmological
constant, but perhaps a dark energy source that varies gradually
as the universe expands, or even a breakdown of general relativity
on large scales.  By itself, however, making dark energy dynamical
does not offer a solution to the coincidence scandal; purely on 
the basis of observations, it seems clear that the universe has
begun to accelerate recently, which implies a scale at which something
new is kicking in.  In particular, it is fruitless to try to 
explain the matter/dark energy coincidence by invoking
mechanisms which make the dark energy density time-dependent in such a
way as to {\it always} be proportional to that in matter.  Such a
scenario would either imply that the dark energy would redshift away
as $\rho_\dark \propto a^{-3}$, which from (\ref{feq}) 
would lead to a non-accelerating universe, or require departures from
conventional general relativity of the type which (as discussed below) 
are excluded by other measurements.  

\section{What might be going on?}

Observations have led us to a picture of the universe which differs dramatically from what we might have expected.  In this section we
discuss possible ways to come to terms with this situation; the approaches
we consider include both attempts to explain a small but nonzero
vacuum energy, and more dramatic ideas which move beyond a 
simple cosmological constant.  We 
certainly are not close to settling on a favored explanation either
for the low value of the vacuum energy nor the recent onset of 
universal acceleration, but we can try to categorize the different
types of conceivable scenarios.

\begin{figure}[t]
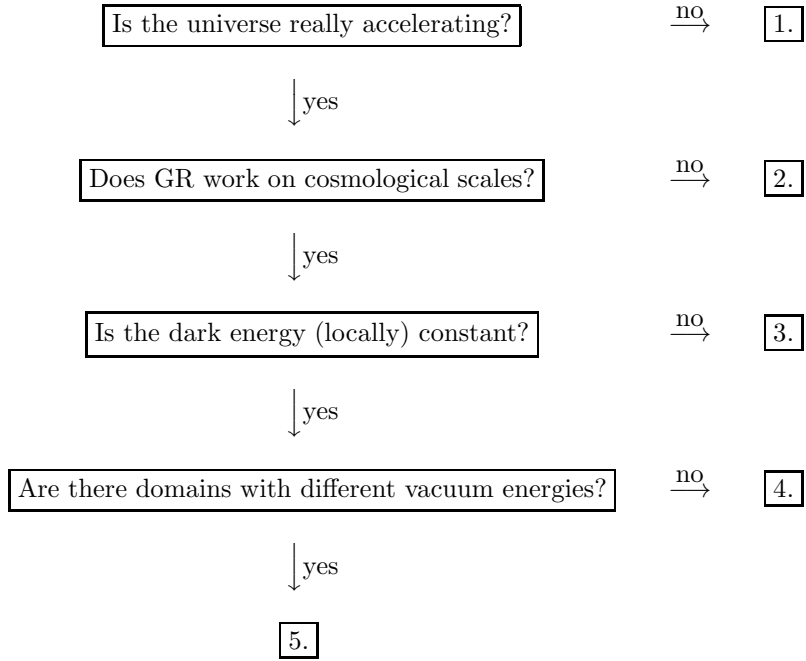

  \[ \hspace{-1cm}
  \begin{array}{ccccc}
  \mbox{\fbox{Is the universe really accelerating?}}  && 
  \stackrel{\mbox{no}}{\longrightarrow} & &
  \fbox{1.}  \cr\cr
  {\Big\downarrow} \mbox{yes} &&&&  \cr\cr
  \mbox{\fbox{Does GR work on cosmological scales?}}
  && \stackrel{\mbox{no}}{\longrightarrow} & &
  \fbox{2.}   \cr\cr
  {\Big\downarrow} \mbox{yes} &&&&  \cr\cr
  \mbox{\fbox{Is the dark energy (locally) constant?}}
  && \stackrel{\mbox{no}}{\longrightarrow} & &
  \fbox{3.}   \cr\cr
  {\Big\downarrow} \mbox{yes} &&&&  \cr\cr
  \mbox{\fbox{Are there domains with different vacuum energies?}}
  && \stackrel{\mbox{no}}{\longrightarrow} && 
  \fbox{4.}   \cr\cr
  {\Big\downarrow} \mbox{yes} &&&&  \cr\cr
  \fbox{5.} ~~~ &&&&
  \end{array}
  \]
  \caption{A flowchart classifying reasons why the universe might
  be accelerating.  The possibilities include:  1. Misinterpretation
  of the data; 2. Breakdown of general relativity; 3. Dynamical
  dark energy; 4. Unique vacuum energy; 5. Environmental selection.}
  \label{flowchart}
\end{figure}

The flowchart portrayed in Figure~\ref{flowchart} represents a
classification of scenarios to explain our observations.  
Depending on the answers to various questions,
we have the following possibilities to explain why the universe
appears to be accelerating:
\begin{enumerate}
\item Misinterpretation of the data.
\item Breakdown of general relativity. 
\item Dynamical dark energy.
\item Unique vacuum energy.
\item Environmental selection.
\end{enumerate}
Let's examine each possibility in turn.

\subsection{Are we misinterpreting the data?}

After the original supernova results (Riess et al.\ 1998, Perlmutter
et al.\ 1999) were announced in
1998, cosmologists converted rather quickly from skepticism about 
universal acceleration to a tentative acceptance, which has grown
substantially stronger with time.  The primary reason for this 
sudden conversion has been the convergence of several complementary
lines of evidence in favor of a concordance model; foremost among
the relevant observations are the anisotropy spectrum of the cosmic
microwave background (Spergel et al.\ 2003) 
and the power spectrum of large-scale
structure (Verde et al.\ 2002), but a number of other methods have yielded
consistent answers.

Nevertheless, it remains conceivable that we have dramatically
misinterpreted the data, and the apparent 
agreement of an $\Omega_\Lambda=0.7$, $\Omega_{\rm M}=0.3$ cosmology with a
variety of observations is masking the true situation.  For example,
the supernova observations rely on the nature of Type Ia supernovae
as ``standardizable candles,'' an empirical fact about low-redshift
supernovae which could somehow fail at high redshifts (although
numerous consistency checks have confirmed basic similarities between
SNe at all redshifts).  Given the many other observations, this failure
would not be enough to invalidate our belief in an accelerating
universe; however, we could further imagine that these other methods
are conspiring to point to the wrong conclusion.  This point of view
has been taken by Blanchard et al.\ (2003), who argue that a 
flat matter-dominated ($\Omega_{\rm M}=1$) universe remains consistent
with the data.  To maintain this idea, it is necessary to discard the
supernova results, to imagine that the Hubble constant is approximately
46~km/sec/Mpc (in contrast to the Key Project determination of
$70\pm7$~km/sec/Mpc, Freedman et al.\ 2001), to interpret data on clusters 
and large-scale structure in a way consistent with $\Omega_{\rm M}=1$,
to relax the conventional assumption that
the power spectrum of density fluctuations can be modeled as a single
power law, and to introduce some source beyond ordinary cold dark
matter (such as massive neutrinos) to suppress power on small 
scales.  To most workers in the field this conspiracy of effects 
seems (even) more unlikely than an accelerating universe.

A yet more drastic route is to imagine that our interpretation of
the observations has been skewed by the usual assumption of an
isotropic universe.  It has been argued (Linde, Linde \&
Mezhlumian 1995)
that some versions of the anthropic principle in an eternally
inflating universe lead to a prediction that most galaxies on
a spacelike hypersurface are actually at the center of spherically
symmetric domains with radially-dependent density distributions;
such a configuration could skew the distance-redshift relation at
large distances even without dark energy.  This picture relies 
heavily on a choice of measure in determining what ``most''
galaxies are like, an issue for which there is no obvious
correct choice.

Finally, we may imagine that there is indeed new physics involved
in making distant supernovae dimmer than we would expect,
but it is the propagation of light which is being altered rather than
the expansion of the universe.  Such a scenario has been worked
out by Csaki, Kaloper \& Terning (2002), who suggested that
photons passing through intergalactic magnetic fields may be
converting into light axion-like particles, resulting in a diminuition
of the total flux from supernovae.  In addition to introducing a new
field with significantly constrained mass and coupling, this model
still requires a form of dark energy to reconcile the flatness of the
universe with the low observed matter density (although the 
required energy source could decay away more rapidly than
ordinary dark energy).

The lengths to which it seems necessary to go in order to avoid
concluding that the universe is accelerating is a strong
argument in favor of the concordance model.

\subsection{Is general relativity breaking down?}

If we believe that we live in a universe which is
homogeneous, isotropic, and accelerating, general relativity (GR)
is unambiguous about the need for some sort of dark energy source.
GR has been fantastically successful in passing classic experimental
tests in the solar system, as well as at predicting the amount
of gravitational radiation emitted from the binary pulsar
(Will 2001).  Nevertheless, the possibility remains open that
gravitation might deviate from conventional GR on scales 
corresponding to the radius of the entire universe.
For our present purposes, such deviations may either be
relevant to the cosmological constant problem, or to the dark
energy puzzle.

The idea behind modifying gravity to address the cosmological 
constant problem is to somehow allow for the vacuum energy to be
large, but yet not lead to an appreciable spacetime curvature
(as manifested in a rapidly expanding universe).  Of course we
still need to allow ordinary matter to warp spacetime, so 
there has to be something special about vacuum energy.  One
special thing is that vacuum energy comes with a negative
pressure $p_{\rm vac}=-\rho_{\rm vac}$, 
as in (\ref{pemr}).  We might therefore imagine a theory
which gave rise to a modified version of the Friedmann equation,
of the form
\be
  H^2 \sim \rho + p\ .
  \label{rhoplusp}
\ee
With such an equation, ordinary matter (for which $p$ vanishes)
leads to conventional expansion, while vacuum energy 
decouples entirely.  Such a theory has been studied 
(Carroll \& Mersini 2001), and may even arise in ``self-tuning''
models of extra dimensions (Arkani-Hamed, Dimopoulos, Kaloper \&
Sundrum, 2000, Kachru, Schulz \& Silverstein 2000).  
Unfortunately, close examination
of self-tuning models reveals that there is a hidden fine-tuning,
expressed as a boundary condition chosen at a naked singularity in
the extra dimension.  Furthermore, any alternative to the conventional
Friedmann equation is also constrained by
observations: any alternative must predict the right abundances
of light elements from Big Bang nucleosynthesis (BBN; see
Burles, Nollett, \& Turner 2001), the correct 
evolution of a sensible spectrum of primordial density fluctuations
into the observed spectrum of temperature anisotropies in the 
Cosmic Microwave Background and the power spectrum of large-scale
structure (Tegmark 2002, Zahn \& Zaldarriaga 2003, Lue, Scoccimarro
\& Starkman 2003),
and that the age of the universe is approximately thirteen
billion years.
The most straightforward test comes from BBN (Carroll \& Kaplinghat
2002, Masso \& Rota 2003), 
since the light-element abundances depend on the expansion rate
during a relatively brief period (rather than on the behavior of
perturbations, or an an integral of the expansion rate over a long
period).  Studies of BBN in alternate
cosmologies indicate that it is possible for modifications of
GR to remain consistent with observations, but only for a very 
narrow set of possibilities.  It seems likely that the 
success of conventional BBN, including its agreement with the baryon
density as determined by CMB fluctuations (Spergel et al.\ 2003),
is not a misleading accident, but rather an indication that 
GR provides an accurate description of cosmology when the universe
was of the order of one minute old.  The idea of modifying GR to
solve the cosmological constant problem is not completely dead,
but is evidently not promising.

Rather than trying to solve the cosmological constant problem, we
can put aside the issue of why the magnitude of the vacuum energy
is small and focus instead on whether the current period of 
acceleration can be traced to a modification of GR.  A necessary
feature of any such attempt is to include a new scale in the theory,
since acceleration has only begun relatively 
recently.\footnote{One way of characterizing this scale is in terms
of the Hubble parameter when the universe starts accelerating,
$H_0\sim 10^{-18} {\rm ~sec}^{-1}$.
It is interesting in this context to recall the coincidence pointed
out by Milgrom (1983), that dark {\it matter} only becomes important
in galaxies when the acceleration due to gravity dips below a fixed
value, $a_0/c \leq 10^{-18} {\rm ~sec}^{-1}$.  Milgrom himself has
suggested that the explanation for this feature of galactic dynamics
can be explained by replacing dark matter by a modified dynamics, and
it is irresistible to speculate that both dark matter and dark energy
could be replaced by a single (as yet undiscovered) 
modified theory of gravity.  However, hope for this possibility seems to be
gradually becoming more difficult to maintain, as different methods
indicate the existence of gravitational forces which point in 
directions other than where ordinary matter is 
(Van Waerbeke et al.\ 2000, Dalal \& Kochanek 2002,
Kneib et al.\ 2003) -- a phenomenon that
is easy to explain with dark matter, but difficult with modified 
gravity -- and explanations are offered for $a_0/c\sim H_0$ within
conventional cold dark matter (Scott, White, Cohn, \& Pierpaoli
2001, Kaplinghat \& Turner 2002).}
From a purely phenomenological point of view we can imagine
modifying the Friedmann equation (\ref{feq}) so that acceleration
kicks in when either the energy density approaches a certain
value $\rho_*$,
\be
  H^2 = {8\pi G \over 3}\left[\rho 
  + \left({\rho \over \rho_*}\right)^\alpha\right]\ ,
\ee
or when the Hubble parameter approaches a certain value $H_*$,
\be
  H^2 + \left({H \over H_*}\right)^\beta = {8\pi G \over 3}\rho \ .
\ee
The former idea has been suggested by Freese \& Lewis 2002, the latter
by Dvali \& Turner 2003; in both cases we can fit the data for
appropriate choices of the new parameters.  It is possible that 
equations of this type arise in brane-world models with large
extra spatial dimensions; it is less clear whether the appropriate
parameters can be derived.  An even more dramatic mechanism also
takes advantage of extra dimensions, but allows for separate
gravitational dynamics on and off of our brane; in this case
gravity can be four-dimensional {\it below} a certain length
scale (which would obviously have to be very large), and appear
higher-dimensional at large distances (Dvali, Gabadadze \&
Porrati 2000, Deffayet, Dvali, \& Gabadadze 2002,
Arkani-Hamed, Dimopoulos, Dvali \& Gabadadze 2002).  These
scenarios can also make the universe accelerate at late times,
and may even lead to testable deviations from GR in the 
solar system (Dvali, Gruzinov, \& Zaldarriaga 2003; Lue and Starkman
2003).

As an alternative to extra dimensions, we may look for an
ordinary four-dimensional modification of GR.
This would be unusual behavior, as we are used to thinking of
effective field theories as breaking down at high energies and
small length scales, but being completely reliable in the 
opposite regime.  Nevertheless, it is worth exploring whether
a simple phenomenological model can easily accommodate the data.
Einstein's equation can be derived by minimizing an action given
by the spacetime integral of the curvature scalar $R$,
\be
  S = \int d^4x \sqrt{|g|}\, R\ .
\ee
A simple way to modify the theory when the curvature becomes
very small (at late times in the universe) is to simply add a
piece proportional to $1/R$,
\be
  S = \int d^4x\sqrt{|g|}\, \left(R - {\mu^4 \over R}\right)\ ,
\ee
where $\mu$ is a parameter with dimensions of mass (Carroll,
Duvvuri, Trodden and Turner 2003).  It is straightforward to show
that this theory admits accelerating solutions; unfortunately, it
also brings to life a new scalar degree of freedom, which may ruin
the success of GR in the solar system (Chiba 2003).  Investigations
are still ongoing to see whether a simple modification of this 
idea could explain the acceleration of the universe while remaining
consistent with experimental tests; in the meantime, the difficulty
in finding a simple extension of GR that does away with the
cosmological constant provides yet more support for the standard
scenario.

\subsection{Is dark energy dynamical?}

If general relativity is correct, cosmic acceleration implies there 
must be a dark energy density which diminishes relatively slowly
as the universe expands.  This can be seen directly from the Friedmann
equation (\ref{feq}), which implies
\be
 {\dot a}^2 \propto a^2 \rho + {\rm constant}\ .
\ee
From this relation, it is clear that the only way to get
acceleration ($\dot a$ increasing) in an expanding universe
is if $\rho$ falls off more
slowly than $a^{-2}$; neither matter ($\rho_\M \propto a^{-3}$)
nor radiation ($\rho_\R\propto a^{-4}$) will do the trick.
Vacuum energy is, of course, strictly constant; but the data 
are consistent with smoothly-distributed sources of dark energy
that vary slowly with time.

There are good reasons to consider dynamical dark
energy as an alternative to an honest cosmological constant.
First, a dynamical energy density can be evolving slowly to zero,
allowing for a solution to the cosmological constant problem which
makes the ultimate vacuum energy vanish exactly.  Second, it poses
an interesting and challenging observational problem to study the
evolution of the dark energy, from which we might learn something
about the underlying physical mechanism.  Perhaps most intriguingly,
allowing the dark energy to evolve opens the possibility
of finding a dynamical solution to the coincidence problem, if the
dynamics are such as to trigger a recent takeover by the dark energy
(independently of, or at least for a wide range of, the 
parameters in the theory).  To date this hope has not quite been
met, but dynamical mechanisms at least allow for the possibility 
(unlike a true cosmological constant).

The simplest possibility along these lines
involves the same kind of source
typically invoked in models of inflation in the very early universe:
a scalar field $\phi$ rolling slowly in a potential, sometimes known as
``quintessence'' (Peebles \& Ratra 1998, Ratra \& Peebles 1998,
Wetterich 1998, Frieman, Hill \& Watkins 1992, Frieman, Hill,
Stebbins \&  Waga 1995, Caldwell, Dave \& Steinhardt 1998,
Huey, Wang, Dave, Caldwell \& Steinhardt 1999).\footnote{While
the potential energy of a light scalar field is the most straightforward
candidate for dynamical dark energy, it is by no means the only
possibility.  Other models included tangled topological defects
(Vilenkin 1984, Spergel \& Pen 1996, Battye, Bucher \& Spergel 1999,
Friedland, Murayama \& Perelstein 2003), curved-spacetime
renormalization effects (Sahni \& Habib 1998,
Parker \& Raval 1999),  trans-planckian
vacuum modes (Mersini, Bastero-Gil and Kanti 2001, Bastero-Gil
\& Mersini 2002, Lemoine, Martin \& Uzan 2003), and Chaplygin
gasses (Kamenshchik, Moschella \& Pasquier 2001).}
The energy density of a scalar field is a sum of kinetic, gradient,
and potential energies,
\be
 \rho_\phi = {1\over 2}{\dot\phi}^2 + {1\over 2}(\nabla\phi)^2 + 
 V(\phi)\ .
 \label{rhophi}
\ee
For a homogeneous field ($\nabla\phi \approx 0$),
the equation of motion in an expanding universe is
\be
 \ddot\phi + 3H\dot\phi + {dV \over d\phi} = 0\ .
\ee
If the slope of the potential $V$ is quite flat, we will have
solutions for which $\phi$ is nearly constant throughout space and
only evolving very gradually with time; the energy density in
such a configuration is
\be
  \rho_\phi \approx V(\phi) \approx {\rm constant}\ .
\ee
Thus, a slowly-rolling scalar field is an appropriate candidate
for dark energy.

However, introducing dynamics opens up the possibility
of introducing new problems, the form and severity
of which will depend on the specific
kind of model being considered.  Most quintessence
models feature scalar fields $\phi$ with masses of order the 
current Hubble scale,
\be
  m_\phi \sim H_0 \sim 10^{-33} {\rm ~eV}\ .
\ee
(Fields with larger masses would typically have already rolled
to the minimum of their potentials.)
In quantum field theory, light scalar fields are
unnatural; renormalization effects tend to drive scalar masses
up to the scale of new physics.  The well-known hierarchy
problem of particle physics amounts to asking why the Higgs
mass, thought to be of order $10^{11}$~eV, should be so much
smaller than the grand unification/Planck scale, 
$10^{25}$-$10^{27}$~eV.  Masses of $10^{-33}$~eV are 
correspondingly harder to understand.  (Strategies toward
understanding include approximate global symmetries,
discussed in section~\ref{direct}, and large kinetic-term
renormalizations, as suggested by Dimopulos \& Thomas 2003.)

Nevertheless, this apparent fine-tuning might be worth the
price, if we were somehow able to explain the coincidence problem.
To date, many investigations have considered scalar fields with
potentials that asymptote gradually to zero, of the form
$e^{1/\phi}$ or $1/\phi$.  These can have cosmologically interesting
properties, including ``tracking'' behavior that makes the current
energy density largely independent of the initial conditions
(Zlatev, Wang \& Steinhardt 1999).  They do not, however,
provide a solution to the coincidence problem, as the era in which
the scalar field begins to dominate is still set by finely-tuned
parameters in the theory.  One way to address the coincidence
problem is to take advantage of the fact that matter/radiation 
equality was a relatively recent occurrence (at least on a 
logarithmic scale); if a scalar field has dynamics which are
sensitive to the difference between matter- and radiation-dominated
universes, we might hope that its energy density becomes constant
only after matter/radiation equality.  An approach which takes this
route is $k$-essence (Armendariz-Picon, Mukhanov \& Steinhardt 2000), 
which modifies the form of the kinetic energy for the scalar field.  
Instead of a conventional kinetic energy $K={1\over 2}(\dot\phi)^2$,
in $k$-essence we posit a form
\be
  K = f(\phi) g(\dot\phi^2)\ ,
\ee
where $f$ and $g$ are functions specified by the model.  
For certain choices of these functions, 
the $k$-essence field naturally tracks the evolution of
the total radiation energy density during radiation domination,
but switches to being almost constant once matter begins to
dominate.  Unfortunately,
it seems necessary to choose a finely-tuned kinetic term to get
the desired behavior (Malquarti, Copeland, \& Liddle 2003).

An alternative possibility is that there is nothing special about
the present era; rather, acceleration is just something that
happens from time to time.  This can be accomplished by oscillating
dark energy (Dodelson, Kaplinghat \& Stewart 2000).
In these models the potential takes the form of a decaying 
exponential (which by itself would give scaling behavior, so that
the dark energy remained proportional to the background density) with
small perturbations superimposed:
\be
  V(\phi) = e^{-\phi}[1 + \alpha\cos(\phi)]\ .
\ee
On average, the dark energy in such a model will track that of
the dominant matter/radiation component; however, there will be
gradual oscillations from a negligible density to a dominant
density and back, on a timescale set by the Hubble parameter,
leading to occasional periods of acceleration.
In the previous section we mentioned the success of the conventional
picture in describing primordial nucleosynthesis (when the scale
factor was $a_{\rm BBN}\sim 10^{-9}$) and temperature
fluctuations imprinted on the CMB at recombination
($a_{\rm CMB}\sim 10^{-3}$), 
which implies that the oscillating scalar
must have had a negligible density during those periods; but 
explicit models are able to accommodate this constraint.
Unfortunately, in neither
the $k$-essence models nor the oscillating models do we have a
compelling particle-physics motivation for the chosen dynamics,
and in both cases the behavior still depends sensitively on the
precise form of parameters and interactions chosen.  Nevertheless,
these theories stand as interesting attempts to address the 
coincidence problem by dynamical means.

\subsection{Did we just get lucky?}

By far the most straightforward explanation for the observed
acceleration of the universe is an absolutely constant vacuum
energy, or cosmological constant.  Even in this case we can 
distinguish between two very different scenarios: one in which 
the vacuum energy is some fixed number that as yet we simply
don't know how to calculate, and an alternative in which there
are many distinct domains in the universe, with different
values of the vacuum energy in each.  In this section we concentrate
on the first possibility.  Note that such a scenario requires that
we essentially give up on finding a dynamical resolution to the
coincidence scandal; instead, the vacuum energy is fixed once and
for all, and we are simply fortunate that it takes on a sufficiently
gentle value that life has enough time and space to exist.

To date, there are not any especially promising approaches to
calculating the vacuum energy and getting the right answer;
it is nevertheless instructive to consider the example of
supersymmetry, which relates to the cosmological constant problem
in an interesting way.
Supersymmetry posits that for each fermionic degree of freedom 
there is a matching bosonic degree of freedom, and vice-versa.
By ``matching'' we mean, for example, that the spin-1/2 electron
must be accompanied by a spin-0 ``selectron'' with the same mass
and charge.  The good news is that, while bosonic
fields contribute a positive vacuum energy, for fermions the
contribution is negative.  Hence, if degrees of freedom exactly
match, the net vacuum energy sums to zero.  Supersymmetry is thus
an example of a theory, other than gravity, where the absolute
zero-point of energy is a meaningful concept.  (This can be traced
to the fact that supersymmetry is a spacetime symmetry, relating
particles of different spins.)

We do not, however, live in a supersymmetric state; there is no
selectron with the same mass and charge as an electron, or we would
have noticed it long ago.  If supersymmetry exists in nature, it must
be broken at some scale $M_\susy$.  In a theory with broken supersymmetry,
the vacuum energy is not expected to vanish, but to be of order
\be
  M_\vac \sim M_\susy\ ,\qquad \qquad \quad{\rm (theory)}
\ee
with $\rho_\vac = M_\vac^4$.  What should $M_\susy$ be?  One nice
feature of supersymmetry is that it helps us understand the 
hierarchy problem -- why the scale of electroweak symmetry breaking
is so much smaller than the scales of quantum gravity or grand
unification.  For supersymmetry to be relevant to the hierarchy
problem, we need the 
supersymmetry-breaking scale to be just above the electroweak 
scale, or
\be
  M_\susy \sim 10^3{\rm ~GeV}\ .
\ee
In fact, this is very close to the experimental bound, and there
is good reason to believe that supersymmetry will be discovered 
soon at Fermilab or CERN, if it is connected to electroweak
physics.  

Unfortunately, we are left with a sizable discrepancy between
theory and observation:
\be
  M_\vac^{\rm (obs)} \sim 10^{-15}M_\susy\ . \qquad {\rm (experiment)}
  \label{mvac2}
\ee
Compared to (\ref{mvac1}), we find that supersymmetry has,
in some sense, solved the problem halfway (on a logarithmic scale).
This is encouraging, as it at least represents a step in the right
direction.  Unfortunately, it is ultimately discouraging, since
(\ref{mvac1}) was simply a guess, while (\ref{mvac2}) is actually
a reliable result in this context; supersymmetry renders the
vacuum energy finite and calculable, but the answer is still
far away from what we need.  (Subtleties in supergravity and string
theory allow us to add a negative contribution to the vacuum
energy, with which we could conceivably tune the answer to zero
or some other small number; but there is no reason for this
tuning to actually happen.)

But perhaps there is something deep about supersymmetry which
we don't understand, and our estimate $M_\vac \sim M_\susy$ is
simply incorrect.  What if instead the correct formula were
\be
  M_\vac \sim \left({M_\susy \over M_{\rm P}}\right) M_\susy\ ?
\ee
In other words, we are guessing that the supersymmetry-breaking
scale is actually the geometric mean of the vacuum scale and
the Planck scale.
Because $M_{\rm P}$ is fifteen orders of magnitude larger
than $M_\susy$, and $M_\susy$ is fifteen orders of magnitude larger
than $M_\vac$, this guess gives us the correct answer!  Unfortunately
this is simply optimistic numerology; there is no theory that
actually yields this answer (although there are speculations in
this direction; Banks 2003).  Still, the simplicity with which we
can write down the formula allows us to dream that an improved
understanding of supersymmetry might eventually yield the 
correct result.

Besides supersymmetry, we do know of other phenomena which may in
principle affect our understanding of vacuum energy.  One 
example is the idea
of large extra dimensions of space, which become possible if the
particles of the Standard Model are confined to a three-dimensional
brane (Arkani-Hamed, Dimopoulos \& Dvali 1998, Randall \& 
Sundrum 1999).  In this case gravity is not simply described by
four-dimensional general relativity, as alluded to in the previous
section.  Furthermore, current experimental bounds 
on simple extra-dimensional models limit the scale characterizing
the extra dimensions to less than $10^{-2}$~cm, which corresponds to
an energy of approximately $10^{-3}$~eV; this is coincidentally the
same as the vacuum-energy scale (\ref{vescale}).  As before, 
nobody has a solid reason why these two
scales should be related, but it is worth searching for one.
The fact that we are forced to take such slim hopes seriously is a
measure of how difficult the cosmological constant problem really
is.

\subsection{Are we witnessing environmental selection?}

If the vacuum energy can in principle be calculated in terms of
other measurable quantities, then we clearly don't yet know how
to calculate it.  Alternatively, however, it may be that the vacuum
energy is not a fundamental quantity, but simply our feature of our
local environment.  We don't turn to fundamental theory for an
explanation of the average temperature of the Earth's atmosphere,
nor are we surprised that this temperature is noticeably larger
than in most places in the universe; perhaps the cosmological
constant is on the same footing.  

To make this idea work, we need
to imagine that there are many different regions of the universe
in which the vacuum energy takes on different values; then we would
expect to find ourselves in a region which was hospitable to our
own existence.  Although most humans don't think of the
vacuum energy as playing any role in their lives, a substantially
larger value than we presently observe would either have led to
a rapid recollapse of the universe (if $\rho_\vac$ were negative)
or an inability to form galaxies (if $\rho_\vac$ were positive).
Depending on the distribution of possible values of $\rho_\vac$,
one can argue that the observed value is in excellent
agreement with what we should expect (Weinberg 1987, Linde
1987, Vilenkin 1995, Efstathiou 1995,
Martel, Shapiro \& Weinberg 1998, Garriga \& Vilenkin 2000, 2003).

The idea of understanding the vacuum energy as a consequence of
environmental selection often goes under the name of the ``anthropic
principle,'' and has an unsavory reputation in some circles.  There are
many bad reasons to be skeptical of this approach, and at least one good
reason.  The bad reasons generally center around the idea that it
is somehow an abrogation of our scientific responsibilities to give
up on calculating something as fundamental as the vacuum energy,
or that the existence of many unseen domains in the universe is a
metaphysical construct without any testable consequences, and
hence unscientific.  The problem with these objections is that they
say nothing about whether environmental selection actually happens;
they are only declarations that we hope it doesn't happen, or it would
be difficult for us to prove once and for all that it does.  The good
reason to be skeptical is that environmental selection only works
under certain special circumstances, and we are far from 
understanding whether those conditions hold in our
universe.  In particular, we need to show that there can be a
huge number of different domains with slightly different values
of the vacuum energy, and that the domains can be big enough that
our entire observable universe is a single domain, and that the
possible variation of other physical quantities from domain to domain
is consistent with what we observe in ours.\footnote{For example,
if we have a theory that allows for any possible value of the
vacuum energy, but insists that the vacuum energy scale be equal
to the supersymmetry breaking scale in each separate domain, 
we haven't solved any problems.}

Recent work in string theory has lent some support to the idea
that there are a wide variety of possible vacuum states rather
than a unique one (Dasgupta, Rajesh \& Sethi 1999,
Bousso \& Polchinski 2000, Feng, March-Russell, Sethi \& Wilczek
2001, Giddings, Kachru \&
Polchinski 2002, Kachru, Kallosh, Linde \& Trivedi 2003, Susskind 
2003, Douglas 2003, Ashok \& Douglas 2003).  String
theorists have been investigating novel ways to compactify
extra dimensions, in which crucial roles are played by
branes and gauge fields.  By taking
different combinations of extra-dimensional geometries, brane
configurations, and gauge-field fluxes, it seems plausible that a
wide variety of states may be constructed, with different local
values of the vacuum energy and other physical parameters.
(The set of configurations is sometimes known as the ``landscape,''
and the discrete set of vacuum configurations is unfortunately known
as the ``discretuum.'')  
An obstacle to understanding these purported solutions is the
role of supersymmetry, which is an important part of string theory
but needs to be broken to obtain a realistic universe.  From the
point of view of a four-dimensional observer, the compactifications
that have small values of the cosmological constant would appear to 
be exactly the states alluded to in the previous section, where
one begins with a supersymmetric state with a negative vacuum energy,
to which supersymmetry breaking adds just the right amount of positive
vacuum energy to give a small overall value.  The necessary
fine-tuning is accomplished simply by imagining that there are
many (more than $10^{100}$) such states, so that even very unlikely
things will sometimes occur.  We still have a long way to go 
before we understand this possibility; in particular, it is not
clear that the many states obtained have all the desired
properties (Banks, Dine \& Motl 2001, Banks, Dine \& Gorbatov 2003), 
or even if they are stable enough to last for the age of the universe
(Hertog, Horowitz \& Maeda, 2003).

Even if such states are allowed, it is necessary to imagine a universe
in which a large number of them actually exist in local regions
widely separated from each other.  As is well known,
inflation works to take a small region
of space and expand it to a size larger than the observable universe;
it is not much of a stretch to imagine that a multitude of different
domains may be separately inflated, each with different vacuum
energies.  Indeed, models of inflation generally tend to be
eternal, in the sense that the universe continues to inflate in
some regions even after inflation has ended in others (Vilenkin 1983,
Linde 1985, Goncharov, Linde \& Mukhanov 1987).
Thus, our observable universe may be separated by inflating regions
from other ``universes'' which have landed in different vacuum
states; this is precisely what is needed to empower the idea
of environmental selection.

Nevertheless, it seems extravagant to imagine a fantastic number
of separate regions of the universe, outside the boundary of
what we can ever possibly observe, just so that we may understand
the value of the vacuum energy in our region.  But again, this
doesn't mean it isn't true.  To decide once and for all will
be extremely difficult, and will at the least require a much
better understanding of how both string theory (or some
alternative) and inflation operate -- an understanding that
we will undoubtedly require a great deal of experimental input
to achieve. 

\section{Observational issues}

From the above discussion, it is clear that theorists are in
desperate need of further input from experiment -- in particular,
we need to know if the dark energy is constant or dynamical, and
if it is dynamical what form it takes.  The observational 
program to test these ideas has been discussed in detail elsewhere
(Sahni \& Starobinski 2000, Carroll 2001,
Peebles \& Ratra 2003); here we briefly draw attention to 
a couple of theoretical issues which can affect
the observational strategies.

\subsection{Equation-of-state parameter}

Given that the universe is accelerating, the next immediate
question is whether the acceleration is caused by a strictly
constant vacuum energy or something else; the obvious place to look
is for some time-dependence to the dark energy density.  In principle any 
behavior is possible, but it is sensible to choose a simple
parameterization which would characterize
dark energy evolution in the measurable regime of relatively
nearby redshifts (order unity or less).  For this purpose it is
common to imagine that the dark energy evolves as a power law
with the scale factor:
\be
  \rho_\dark \propto a^{-n} \ .
  \label{neq}
\ee
Even if $\rho_\dark$ is not strictly a power law, this ansatz
can be a useful characterization of its effective behavior at
low redshifts.  We can then define an equation-of-state
parameter relating the energy density to the pressure,
\be
  p = w \rho\ .
  \label{eos}
\ee
Using the equation of energy-momentum
conservation,
\be
  \dot\rho = -3(\rho + p){{\dot a}\over a}\ ,
\ee
a constant exponent $n$ of (\ref{neq}) implies a constant $w$ with
\be
  n = 3(1+w)\ .
\ee
As $n$ varies from 3 (matter) to 0 (cosmological constant), $w$
varies from $0$ to $-1$.  
\begin{figure}[t]
  \centering
  \includegraphics[height=8cm]{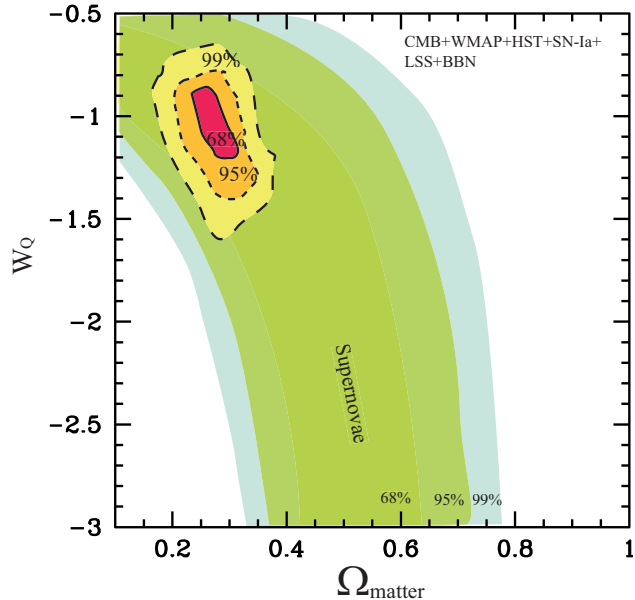}
  \caption{Limits on the equation-of-state parameter $w$ in a flat
  universe, where $\Omega_\M + \Omega_X = 1$. From 
  Melchiorri, Mersini, {\"O}dman \& Trodden 2003.}
  \label{wlimits}
\end{figure}
Some limits from supernovae, large-scale structure, and the CMB from
Melchiorri, Mersini, {\"O}dman \& Trodden 2003
are shown in Figure (\ref{wlimits}); see Spergel et al\ 2003 for
limits from WMAP observations of the cosmic microwave background,
Schuecker et al.\ 2003 for limits from X-ray clusters, 
and Tonry et al.\ 2003 and Knop et al.\ 2003 for more recent
supernova limits.  These constraints
apply to the $\Omega_\M$-$w$ plane, under the assumption that the
universe is flat ($\Omega_\M + \Omega_\dark = 1$).  We see that
the observationally favored region features $\Omega_\M \approx
0.3$ and an honest cosmological constant, $w=-1$.  However, there
is room for alternatives; one of the most important tasks
of observational cosmology will be to reduce the error regions on plots
such as this to pin down precise values of these parameters.

It is clear that $w=-1$ is a special value; for $w>-1$ the
dark energy density slowly decreases as the universe expands,
while for $w<-1$ it would actually be {\it increasing}.  In most
conventional models, unsurprisingly, we have $w\geq -1$; this is
also required (for sources with positive energy densities) by
the energy conditions of general relativity (Garnavich et al.\ 1998).  
Nevertheless, it
is interesting to ask whether we should bother to consider
$w<-1$ (Parker \& Raval 1999, Sahni \& Starobinski 2000, Caldwell
2002, Carroll, Hoffman \& Trodden 2003).  
If $w$ is constant in such a model, the universe will
expand ever faster until a future singularity is reached, the
``Big Rip'' (Caldwell, Kamionkowski \& Weinberg 2003); but such
behavior is by no means necessary.  An explicit model
is given by so-called phantom fields (Caldwell 2002),
scalar fields with negative kinetic and gradient energy, 
\be
  \rho_\phi = -{1\over 2}{\dot\phi}^2 - {1\over 2}(\nabla\phi)^2 + 
  V(\phi)\ ,
\ee
in contrast with the conventional expression (\ref{rhophi}).
(A phantom may be thought of as a physical realization of
the ``ghost'' fields used in some calculations in quantum field 
theory.)  A phantom field rolls to the maximum of its potential,
rather than the minimum; if there is a maximum with positive
potential energy, we will have $w<-1$ while the field is rolling,
but it will settle into a state with $w=-1$.

However, such fields are very dangerous in particle physics; the
excitations of the phantom will be negative-mass particles, and
therefore allow for the decay of empty space into a collection
of positive-energy ordinary particles and negative-energy phantoms.
Naively the decay rate is infinite, because there is no boundary
to the allowed phase space; if we impose a cutoff by 
hand by disallowing momenta greater than $10^{-3}$~eV, the vacuum
can be stable for the age of the universe (Carroll, Hoffman \&
Trodden 2003).  Of course, there may be other ways to get
$w<-1$ other than a simple phantom field (Parker \& Raval
1999, Dvali \& Turner 2003), and there is a lurking danger that
a rapidly time-varying equation of state might trick you into
thinking that $w<-1$ (Maor, Brustein, McMahon, \& Steinhardt, 2002).
The moral of the story should be that theorists proposing models
with $w<-1$ should be very careful to check that their theories
are sufficiently stable, while observers should be open-minded
and include $w<-1$ in the parameter space they constrain.
To say the least, a convincing measurement that the effective
value of $w$ were less than $-1$ would be an important discovery,
the possibility of which one would not want to exclude
{\it a priori}.

\subsection{Direct detection of dark energy}
\label{direct}

If dark energy is dynamical rather than simply a constant,
it is able to interact with other fields, including those of
the Standard Model of particle physics.  For the particular
example of an ultra-light scalar field, interactions introduce
the possibility of two observable phenomena:  long-range
``fifth forces'' and time-dependence of the constants of
nature.  Even if a dark-energy scalar 
$\phi$ interacts with ordinary matter only through
indirect gravitational-strength couplings, searches
for these phenomena
should have already enabled us to detect the quintessence
field (Carroll 1998, Dvali \& Zaldarriaga 2002); 
to avoid detection, we need to introduce
dimensionless suppression factors of order $10^{-5}$ or less
in the coupling constants.  On the other hand, there has been
some evidence from quasar absorption spectra that the fine-structure
constant $\alpha$ was slightly smaller ($\Delta\alpha/\alpha \sim
-10^{-5}$) at redshifts $z\sim 0.5-3$ (Murphy et al.\ 2001).
On the most optimistic reading, this apparent shift might be
direct evidence of a quintessence field; this would place strong
constraints on the quintessence potential (Chiba \& Kohri 2002).
Before such an interpretation is accepted, however, it will 
be necessary to be certain that all possible sources of systematic
error in the quasar measurements are understood, and that models
can be constructed which fits the quasar data while remaining
consistent with other experimental bounds (Uzan 2003).

More likely, we should work to construct particle
physics models of quintessence in which both the mass
and the interactions of the scalar field with ordinary matter
are naturally suppressed.  These requirements are met by
Pseudo-Nambu-Goldstone bosons (PNGB's) 
(Frieman, Hill \& Watkins 1992, Frieman, Hill, 
Stebbins \& Waga 1995), which arise in 
models with approximate global symmetries of the form
\be
  \phi \rightarrow \phi + {\rm constant}.
  \label{pgbsym}
\ee
Clearly such a symmetry should not be exact, or the potential would
be precisely flat; however, even an approximate symmetry can
naturally suppress masses and couplings.  PNGB's typically
arise as the angular degrees of freedom in Mexican-hat
potentials that are ``tilted'' by a small explicitly symmetry
breaking, and the PNGB potential takes on a sinusoidal form:
\be
  V(\phi) = \mu^4[1+ \cos(\phi)]\ .
\ee 
Fields of this type are ubiquitous in string theory, and it is
possible that one of them may have the right properties to be the
dark energy (Choi 2000; Kim 2000; Kim \& Nilles 2003).
Supersymmetric versions have been studied by Bi, Li \& Zhang (2003).
Interestingly, while the symmetry (\ref{pgbsym}) suppresses most
possible interactions with ordinary matter, it
leaves open one possibility -- a pseudoscalar electromagnetic
interaction in which $\phi$ couples to ${\bf E}\cdot {\bf B}$.
The effect of such an interaction would be to gradually rotate
the plane of polarization of light from distant sources
(Carroll 1998, Lue, Wang \& Kamionkowski 1999); current limits
on such a rotation are not quite sensitive enough to tightly
constrain this coupling.  It is therefore very plausible that
a pseudoscalar quintessence field will be directly detected by
improved polarization measurements in the near future.

Even if we manage to avoid detectable interactions between 
dark energy and ordinary matter, we may still consider
the possibility of nontrivial interactions between dark matter
and dark energy.  Numerous models along these lines have been
proposed (Casas, Garcia-Bellido \& Quiros 1992, Wetterich 1995,
Anderson \& Carroll 1998, Amendola 2000, Bean 2001, Comelli,
Pietroni \& Riotto 2003; for recent work and further
references see Farrar \& Peebles 2003, Hoffman 2003, Fardon,
Nelson \& Weiner 2003).
If these two dark components constitute 95\% of the universe, 
the idea that they are separate and non-interacting may simply
be a useful starting point.  Investigations thus far seem to 
indicate that some sorts of interactions are possible, but
constraints imposed by the cosmic microwave background and
large-scale structure are actually able to exclude a wide range
of possibilities.  It may be that the richness of interaction we
observe in the ordinary-matter sector is an exception rather than
the rule.

Finally, our natural tendencies toward economy of explanation
inspires us to consider models in which the dark energy does
more than simply accelerate the universe.  Accordingly, models
have been proposed in which quintessence is involved in
inflation (Peebles \& Vilenkin 1999, Copeland, Liddle and
Lidsey 2001) as well as in baryogenesis (Li, Wang, Feng
\& Zhang 2002, De~Felice, Nasri \& Trodden 2003, Gu, Wang
\& Zhang 2003).  Perhaps
the success or failure of these mechanisms will one day provide
a clue to a more comprehensive picture of cosmology.

\section{Conclusions}

The acceleration of the universe presents us with mysteries and
opportunities.  The fact that this behavior is so puzzling is a
sign that there is something fundamental we don't understand.
We don't even know whether our misunderstanding originates with
gravity as described by general relativity, with some source of
dynamical or constant dark energy, or with the structure of
the universe on ultra-large scales.  Regardless of what the
answer is, we seem poised to discover something profound about
how the universe works.

\section*{Acknowledgments}

It is a pleasure to thank Wendy Freedman for organizing a stimulating 
meeting, and participants at the Seven Pines Symposium on ``The
Concept of the Vacuum in Physics'' and the Kavli Institute for 
Theoretical Physics program on ``String Theory and Cosmology'' for
numerous helpful conversations.  This work was supported in part by
U.S. Dept.\ of Energy contract DE-FG02-90ER-40560, National Science
Foundation Grants PHY01-14422 (CfCP) and PHY99-07949 (KITP),
and the David and Lucile Packard Foundation.


\begin{thereferences}{}

\bibitem{amendola} Amendola, L.\ 2000, \prd 62, 043511 [astro-ph/9908023]

\bibitem{1998cosm.work..227A} Anderson, G.~W.~\& 
Carroll, S.~M.\ 1998, COSMO-97, First International Workshop on Particle 
Physics and the Early Universe, 227 [astro-ph/9711288]

\bibitem{add} Arkani-Hamed, N., Dimopoulos, S., \& Dvali, G.\ 1998,
Phys.\ Lett.\ B429, 263 [hep-ph/9803315]

\bibitem{addg} Arkani-Hamed, N., Dimopoulos, S., Dvali, G., \&
Gabadadze, G.\ 2002, hep-th/0209227 

\bibitem{adks} Arkani-Hamed, N., Dimopoulos, S., Kaloper, N.
\& Sundrum, R.\ 2000,
Phys.\ Lett.\ B480, 193 [hep-th/0001197]

\bibitem{2002PhRvD..66f4008A} Armendariz-Picon, 
C.\ 2002, \prd 66, 64008 

\bibitem{armen}Armendariz-Picon, C., Mukhanov, V., 
\& Steinhardt, P.~J.\ 2000, Phys.\ Rev.\ Lett.\  85, 4438 
[astro-ph/0004134]

\bibitem{ad}
Ashok, S.\ \& Douglas, M.~R.\ 2003, [hep-th/0307049]

\bibitem{banks}  Banks, T.\ 2003, hep-th/0305206

\bibitem{bdg}  Banks, T., Dine, M. and Gorbatov, E.\ 2003, 
hep-th/0309170

\bibitem{bdm} Banks,  T., Dine, M. and Motl, L.\ 2003,
JHEP 0101, 031 [hep-th/0007206]

\bibitem{mar}
Bastero-Gil, M.\ \& Mersini, L.\ 2002, \prd 67, 103519 [hep-th/0205271]

\bibitem{battye}
Battye, R.~A., Bucher, M.\ \& Spergel, D.\ 1999, astro-ph/9908047

\bibitem{bean} Bean, R.\ 2001, \prd 64, 123516 [astro-ph/0104464]

\bibitem{bi}
Bi, X.-J., Li, M., \& Zhang, X.\ 2003, hep-ph/0308218

\bibitem{Blanchard:2003du}
Blanchard, A., Douspis, M., Rowan-Robinson, M. \& Sarkar, S. 2003,
astro-ph/0304237

\bibitem{bp}  Bousso, R. \& Polchinski, J.\ 2000,
JHEP 0006, 006 [hep-th/0004134]

\bibitem{2001ApJ...552L...1B} Burles, S., 
Nollett, K.~M., \& Turner, M.~S.\ 2001, \apj 552, L1 

\bibitem{caldwell}Caldwell, R.~R.\ 2002, 
Phys.\ Lett.\  B 545, 23 
[astro-ph/9908168]

\bibitem{q1}Caldwell, R.~R., Dave, R., \& Steinhardt, P.~J.\ 1998, 
Phys.\ Rev.\ Lett. 80, 1582  [astro-ph/9708069]

\bibitem{rip} Caldwell, R.~R., Kamionkowski, M., \& Weinberg, N.~N.\ 2003, Phys. Rev. Lett. 91, 71301 

\bibitem{world}Carroll, S.~M.\ 1998, Phys. Rev. Lett. 81, 3067 
[astro-ph/9806099]

\bibitem{carroll}Carroll, S.~M.\ 2001, Living 
Reviews in Relativity 4, 1 
[astro-ph/0004075]

\bibitem{Carroll:2003wy}Carroll, 
S.~M., Duvvuri, V., Trodden, M. \& Turner, M.~S.\ 2003,
astro-ph/0306438

\bibitem{Carroll:2003st} Carroll, 
S.~M., Hoffman, M., \& Trodden, M.\ 2003, \prd 68, 23509
[astro-ph/0301273]

\bibitem{ck}Carroll, 
S.~M.~\& Kaplinghat, M.\ 2002, \prd 65, 63507 
[astro-ph/0108002]

\bibitem{carrollmersini}
Carroll, S.~M.~\& 
Mersini, L.\ 2001, \prd 64, 124008

\bibitem{casas} Casas, J.~A., Garcia-Bellido, J.\ \& Quiros, M.\ 
1992, Class.\ Quant.\ Grav.\ 9, 1371

\bibitem{chiba}
Chiba, T.\ 2003, astro-ph/0307338

\bibitem{2002PThPh.107..631C} Chiba, T.~\& Kohri, K.\ 
2002, Progress of Theoretical Physics,107, 631 

\bibitem{choi} Choi, K.\ 2000, \prd 62, 043509 [hep-ph/9902292]

\bibitem{comelli}
Comelli, D., Pietroni, M., \& Riotto, A.\ 2003, hep-ph/0302080

\bibitem{copeland}
Copeland, E.~ J., Liddle, A.~R., \& Lidsey, J.~E.\ 2001, \prd
64, 023509 [astro-ph/0006421]

\bibitem{csaki}
Csaki, C., Kaloper, N., \& Terning, J.\ 2002, Phys.\ Rev.\ Lett.\ 
88,161302 [hep-ph/0111311]

\bibitem{2002ApJ...572...25D} Dalal, N.~\& 
Kochanek, C.~S.\ 2002, \apj 572, 25 

\bibitem{drs}
Dasgupta, K., Rajesh, G.\ \& Sethi, S.\ 1999, JHEP 9908, 023
[hep-th/9908088]

\bibitem{dnt}
De Felice, A., Nasri, S., \& Trodden, M.\ 2003, \prd 67, 043509
[hep-ph/0207211]

\bibitem{ddg} 
Deffayet, C., Dvali, G., \& Gabadadze, G.\ 2002, \prd 65, 44023 
[astro-ph/0105068]

\bibitem{dt}
Dimopoulos, S.\ \& Thomas, S.\ 2003, Phys.\ Lett.\ B 537,
19 [hep-th/0307004]

\bibitem{dks}
Dodelson, S., Kaplinghat, M., \& Stewart, E.\ 2000, 
Phys.\ Rev.\ Lett.\ ,85, 5276 
[astro-ph/0002360]

\bibitem{douglas}
Douglas, M.~R.\ 2003, JHEP 0305, 046 [hep-th/0303194]

\bibitem{dgp}  Dvali, G., Gabadadze, G.\ \& Porrati, M.\ 2000,
Phys.\ Lett.\ B 485, 208 [hep-th/0005016 ]

\bibitem{2003PhRvD..68b4012D} Dvali, 
G., Gruzinov, A., \& Zaldarriaga, M.\ 2003, \prd 68, 24012 

\bibitem{mfe} 
Dvali, G., \& Turner, M.~S.\ 2003, astro-ph/0301510 

\bibitem{dz} Dvali, G.~\& 
Zaldarriaga, M.\ 2002, Phys.\ Rev.\ Lett.\  88, 91303

\bibitem{efstathiou}
Efstathiou, G. 1995, \mnras 274, L73

\bibitem{fardon}
Fardon, R., Nelson, A.~N., \& Weiner, N.\ 2003, astro-ph/0309800

\bibitem{fp} Farrar, G.~R.\  \& Peebles, P.~J.~E.\ 2003,
astro-ph/0104464

\bibitem{salt}
Feng, J.~L., March-Russell, J., Sethi, S.\ \& Wilczek, F.\ 2001,
Nucl.\ Phys.\ B 602, 307 [hep-th/0005276]

\bibitem{freedman} Freedman, W.\ et al.\ 2001,
\apj 553, 47

\bibitem{card}  Freese, K.\ \& Lewis, M.\ 2002,
Phys. Lett. B540, 1 [astro-ph/0201229]

\bibitem{fmp}
Friedland, A., Murayama, H., \& Perelstein, M.\ 2003,
\prd 67, 043519 [astro-ph/0205520]

\bibitem{pngb1}Frieman, 
J.~A., Hill, C.~T., \& Watkins, R.\ 1992, \prd 46, 1226 

\bibitem{pngb2}
Frieman, J.~A., Hill, C.~T., Stebbins, A., \& Waga, I.\ 1995, 
Phys. Rev. Lett. 75, 2077 
 [astro-ph/9505060]

\bibitem{gb} Garcia-Bellido, J.\ 1993, Int.\ J.\ Mod.\ Phys.\ 
D2, 85

\bibitem{garnavich}Garnavich, P.~M.~et 
al.\ 1998, \apj 509, 74 
[astro-ph/9806396]

\bibitem{gv}
Garriga, J.~\& 
Vilenkin, A.\ 2000, \prd 61, 83502 
[astro-ph/9908115]

\bibitem{gv2}Garriga, J.~\& 
Vilenkin, A.\ 2003, \prd 67, 43503 

\bibitem{gkp} Giddings, S.~B., Kachru, S. \& Polchinski, J.\ 2002,
\prd 66, 106006 [hep-th/0105097]

\bibitem{goncharov} Goncharov, A.~S., Linde, A.~D., \& Mukhanov, V.~F.\
1987, Int.\ J.\ Mod.\ Phys.\ A2, 561

\bibitem{gu}
Gu, P., Wang, X., \& Zhang, X.\ 2003, hep-ph/0307148

\bibitem{hhm} Hertog, T., Horowitz, G.~T., Maeda, K.\ 2003,
JHEP 0305, 060 [hep-th/0304199]

\bibitem{hoffman} Hoffman, M.\ 2003, astro-ph/0307350

\bibitem{q2}Huey, G., Wang, L., Dave, 
R., Caldwell, R.~R., \& Steinhardt, P.~J.\ 1999, \prd 59, 63005 
 [astro-ph/9804285]

\bibitem{kklt}  Kachru, S., Kallosh, R., Linde, A., \& Trivedi, S.~P.\
2003, hep-th/0301240

\bibitem{kss} Kachru, S., Schulz, M.~B., \& Silverstein, E.\ 2000,
\prd 62, 045021 [hep-th/0001206]

\bibitem{chap}
Kamenschchik, A., Moschella, U., \& Pasquier, V., 2001,
Phys.\ Lett.\ B 511, 265.

\bibitem{kt}
Kaplinghat, M. \& Turner, M.~S.\ 2002,
\apj 569, L19 
[astro-ph/0107284]

\bibitem{kim} Kim, J.~E.\ 2000, JHEP 0006, 016 [hep-ph/9907528]

\bibitem{kimnil}Kim, J.~E.\ \& Nilles, H.~P..\ 2003,
Phys.\ Lett.\ B553, 1 [hep-ph/0210402]

\bibitem{kneib} Kneib, J.-P., Hudelot, P., Ellis, R.~S., Treu, T.,
Smith, G.~P., Marshall, P., Czoske, O., Smail, I.\ \& Natarajan, P.\ 2003,
astro-ph/0307299.

\bibitem{Knop:2003iy}
Knop, R.~A. et al.\ 2003,
astro-ph/0309368

\bibitem{lemoine}
Lemoine, M., Martin, J.\ \& Uzan, J.-P., 2003, \prd 67,
103520 [hep-th/021202]

\bibitem{li}
Li, M., Wang, X., Feng, B., \& Zhang, X.\ 2002, \prd 65, 103511
[hep-ph/0112069]

\bibitem{linde1} Linde, A.\ 1986, Phys.\ Lett.\ B175, 395

\bibitem{linde2} 
Linde, A.\ 1987, in {\sl 300 Years of Gravitation}, ed. by S.~W.\ Hawking
and W.\ Israel (Cambridge: Cambridge University Press)

\bibitem{Linde:1994gy}
Linde, A.~D., Linde, D.~A. \& Mezhlumian, A. 1995,
Phys.\ Lett.\ B 345, 203
[hep-th/9411111]

\bibitem{lss} Lue, A., Scoccimarro, R.\ \& Starkman, G.\ 2003,
astro-ph/0307034

\bibitem{luestark} Lue, A.\ \& Starkman, G.\ 2003,
\prd 67, 064002 [astro-ph/0212083]

\bibitem{lwk} Lue, A., Wang, L.-M., \& Kamionkowski, M.\ 1999,
Phys.\ Rev.\ Lett.\ 83, 1506 [astro-ph/9812088]

\bibitem{2003PhRvD..68b3512M} 
Malquarti, M., Copeland, E.~J., \& Liddle, A.~R.\ 2003, \prd 68, 23512

\bibitem{2002PhRvD..65l3003M} 
Maor, I., Brustein, R., McMahon, J., \& Steinhardt, P.~J.\ 2002, \prd 65, 
123003 

\bibitem{msw}
Martel, H., Shapiro, P.~R.\ \& Weinberg, S.\ 1998
\apj 492,
29 [astro-ph/9701099]

\bibitem{mr} Masso, E.\ \& Rota, F.\ 2003, astro-ph/0302554

\bibitem{Melchiorri:2002ux} Melchiorri, A., Mersini, L., {\" 
O}dman, C.~J., \& Trodden, M.\ 2003, \prd 68, 43509 
[astro-ph/0211522]

\bibitem{mersini}
Mersini, L., Bastero-Gil, M., \& Kanti, P.\ 2001, \prd 64, 043508
[hep-ph/0101210]

\bibitem{milgrom}
Milgrom, M.\ 1983, \apj  270, 365

\bibitem{2001MNRAS.327.1208M} Murphy, M.~T., Webb, 
J.~K., Flambaum, V.~V., Dzuba, V.~A., Churchill, C.~W., Prochaska, J.~X., 
Barrow, J.~D., \& Wolfe, A.~M.\ 2001, \mnras 327, 1208 

\bibitem{pad}
Padmanabhan, T.\ 2003, Physics Reports  380, 235 [hep-th/0212290]

\bibitem{parker}
Parker, L.\ \& Raval., A.\ 1999, \prd 60, 063512
[gr-qc/9905031]

\bibitem{pr1}Peebles, P.~J.~\& 
Ratra, B.\ 1988, \apj  325, L17

\bibitem{Peebles:2002gy} Peebles, P.~J.~E.\& 
Ratra, B.\ 2003, Reviews of Modern Physics 75, 559

\bibitem{pv}
Peebles, P.~J.~E.\& Vilenkin, A.., 1999 \prd 59, 063505
[astro-ph/9810509]

\bibitem{perl} Perlmutter, S.~et 
al.\ 1999, \apj 517, 565
[astro-ph/9812133]

\bibitem{rs2}  Randall, L. \& Sundrum, R. 1999,
Phys.\ Rev.\ Lett.\ 83, 4690 [hep-th/9906064]

\bibitem{pr2} Ratra. B. \& Peebles, P.~J.\ 1988 \prd  37 , 3406 

\bibitem{riess}
Riess, A.~G.~et al.\ 
1998, \aj 116, 1009 
[astro-ph/9805201]

\bibitem{sh}
Sahni, V.\ \& Habib, S.\ 1998, Phys.\ Rev.\ Lett.\  81, 1766
[hep-ph/9808204]

\bibitem{sahni} Sahni, V.~\& 
Starobinsky, A.\ 2000, International Journal of Modern Physics D 9, 373 [astro-ph/9904389]

\bibitem{schuecker}
Schuecker, P., Caldwell, R.~R., B\"ohringer, H., Collins, C.~A., Guzzo, L., \&
Weinberg, N.~N.\  2003, \aa 402, 53 [astro-ph/0211480]

\bibitem{swcp} Scott, D., White, M., Cohn, J.~D., \& Pierpaoli, E.\ 2001,
astro-ph/0104435

\bibitem{spergelpen}
Spergel, D.~N.\ \& Pen, U.-L.\ 1997, \apj 491, L67 [astro-ph/9611198]

\bibitem{wmap}  Spergel, D.~N.~et al.\ 
2003, \apjs 148, 175

\bibitem{sussland}  Susskind, L.\ 2003, hep-th/0302219

\bibitem{tegmark02} Tegmark, M.\ 2002, \prd 66, 
103507 [astro-ph/0101354]

\bibitem{Tonry:2003zg}
Tonry, J.~L.~et al.\ 2003,
\apj  594, 1 
[astro-ph/0305008]

\bibitem{2003RvMP...75..403U} Uzan, J.\ 2003, Reviews of Modern 
Physics 75, 403 

\bibitem{2000A&A...358...30V} Van Waerbeke, 
L.~et al.\ 2000, \aa 358, 30 

\bibitem{2002MNRAS.335..432V} Verde, L.~et al.\ 2002, 
\mnras 335, 432 

\bibitem{vilenkini}Vilenkin, A.\ 1983, \prd 27, 2848

\bibitem{vilenkin3}Vilenkin, A.\ 1984, 
Phys.\ Rev.\ Lett.\ 53, 1016


\bibitem{vilenkin}Vilenkin, A.\ 1995, 
Phys.\ Rev.\ Lett.\ 74, 846  [gr-qc/9406010]

\bibitem{weinberga}
Weinberg, S.\ 1987, Phys.\ Rev.\ Lett.\   59, 2607 

\bibitem{weinberg}
Weinberg, S. 1989,
Rev.\ Mod.\ Phys.\ 61, 1 

\bibitem{wett2} Wetterich, C.\ 1995, \aa 301, 321

\bibitem{wett} Wetterich, C.\ 1998, Nucl.\ Phys.\ B302, 668

\bibitem{will} Will, C.~M.\ 2001, Living 
Reviews in Relativity 4, 4
[gr-qc/0103036]

\bibitem{zz} Zahn, O.\ \& Zaldarriaga, M.\ 2003,
\prd 67, 0630002 [astro-ph/0212360]

\bibitem{zlatev}
Zlatev, I, Wang, L. \& Steinhardt, P.~J.
Phys.\ Rev.\ Lett.\  {82}, 896 
[astro-ph/9807002]

\end{thereferences}
\end{document}